\documentclass{article}

\usepackage[preprint]{neurips_2021}

\usepackage[utf8]{inputenc} 
\usepackage[T1]{fontenc}    
\usepackage{hyperref}       
\usepackage{url}            
\usepackage{booktabs}       
\usepackage{amsfonts}       
\usepackage{nicefrac}       
\usepackage{microtype}      
\usepackage{xcolor}         

\title{Can Machine Learning be Moral?}

\author{%
  Miguel Sicart \\
  Digital Design Department\\
  IT University of Copenhagen\\
  Denmark \\
  \texttt{miguel@itu.dk} \\
 \And
  Irina Shklovski \\
  Department of Computer Science\\
  University of Copenhagen\\
  Denmark \\
  \texttt{ias@di.ku.dk}
 \AND
 Mirabelle Jones \\
  Department of Computer Science\\
  University of Copenhagen\\
  Denmark \\
  \texttt{msd@di.ku.dk}
}

\begin{document}

\maketitle

\section{Introduction}
The ethics of Machine Learning has become an unavoidable topic in the AI Community. The deployment of machine learning systems in multiple social contexts has resulted in a closer ethical scrutiny of the design, development, and application of these systems. The AI/ML community has come to terms with the imperative to think about the ethical implications of machine learning, not only as a product but also as a practice (Birhane, 2021; Shen et al. 2021). The critical question that is troubling many debates is what can constitute an ethically accountable machine learning system. In this paper we explore possibilities for ethical evaluation of machine learning methodologies. We scrutinize techniques, methods and technical practices in machine learning from a relational ethics perspective, taking into consideration how machine learning systems are part of the world and how they relate to different forms of agency. Taking a page from Phil Agre (1997) we use the notion of a critical technical practice as a means of analysis of machine learning approaches. Our radical proposal is that supervised learning appears to be the only machine learning method that is ethically defensible. 

\section{Machine Learning and its Ethical Discontent}
Issues like bias and the moral challenges of data gathering and classification are central to the problems of developing and deploying ethically accountable machine learning systems (Greene et al. 2019). Where data and bias emerged as the primary movers to bring attention to the potential ethical foibles of machine learning, we know that these are not the whole story. In fact problem formulation, model choice, inductive bias and misapplication of machine learning techniques account for a significant proportion of the problems that emerge (Gibney, 2021). These, however, are tied together with available data and the choices made given the data. As a result, each of the major machine learning techniques presents variations of ethical problems around bias and data classification. In supervised learning, the ethical challenges reside with the bias in the data as well as with the labelling of training data. In unsupervised learning, the problem resides with the bias in the data as well as with the lack of human oversight during the training process. In reinforcement learning, the problem lies with the modelling of the rewards and the environment as presented to the agent, as well as with the definition of the different possible responses of the agents to situations in the environment. 

Many agree that the most obvious and perhaps best tractable ethical issues with machine learning involve the origin, provenance and inherent issues of the available data and to a large extent the curation required in preparing data sets for training models (Mehrabi et al, 2021). The latter issue is interesting because part of the problem with curation and labeling data emerges from the fact that all humans are biased in myriad ways and our judgements can be erroneous (Barbosa `I\&' Chen, 2019. By removing or minimizing the role of human agency in the training process, reinforcement and unsupervised learning have tried to address this fact. By removing humans from the process, these approaches are trusting in an engineering solution that assumes that formalized models are by definition objective and ethically neutral. However, removing human agency from the training phase also removes any possibility for ethical accountability and oversight from the critical part of machine learning. In response we see a series of human-in-the-loop solutions that have tried to combine the engineering benefits of these approaches with the accountability of human agency (Gupta et al. 2020, Sloane et al. 2020). These include fairness analytics tools such as AIFairness360, the What-If tool, and explainability tools such as LIME. However, these practices only insert humans at predetermined locations in the process, and do not specify the type of moral action and accountability that the human should add to the loop. In other words, the human may supervise the loop, but there are limits to what can be questioned and how. With these arguments in mind, we consider what it takes for an ML system and technical practice to become ethically accountable.

\section{Supervised Learning as the Only Ethical Machine Learning}
We argue that supervised learning is the only ethical machine learning approach. As supervised learning requires active human intervention in understanding the data and creating the training labels, it is the Machine Learning technique that allows for active  reflection and engagement with the technical practice of model training. 

Categorizing data to train a model is a moral act. It is imperative then that those doing the categorizing or at least those setting up the conditions for categorizing to be done understand that these categories not only may reproduce bias, but also that they are applied to probably biased data. Supervised learning opens up for several positions of ethical engagement with a machine learning process: in the data selection, the data classification, and the reviewing of the accuracy of the results produced by the model. Each of these steps offers opportunities for interrogation and reflection.

We propose to frame Supervised Learning as a type of Critical Technical Practice that defines the act of labeling and classifying data sets for machine learning models as an ethical process. Following Agre’s mandate of reforming AI, we propose to locate the source of accountability not with the algorithms or the data sets, but with the relations between the people that engage in labeling the data and the machine learning system that will be used to train a model. Similar to other forms of learning, we situate the ethical responsibility and accountability with the teacher, not with the students or the physical classroom. Sure the students are responsible for their own learning, but it is the teacher's ethical responsibility to create the right conditions for learning. 

\section{Implications for Machine Learning}
Our argument has these implications:
\begin{itemize}
    \item Any type of ML approach that does not demand ethical engagement with both the data and its labelling will be inevitably ethically flawed.
    \item Human-in-the-loop approaches are not addressing ethical issues in ML since human ethical agency does not have an effect in the loop. Our approach, arguing that data modelling and classification are inherently moral activities that should be central to ML, does address this issue.
    \item Supervised Learning as the only moral type of ML implies that the production of ML models will become more laborious, as the usage of ethically questionable methods for mass labelling of data like Mechanical Turk would undermine the critical technical practice of ML. There is no ethical ML without ethical labor.
\end{itemize}

The next stage of our research is to specify the requirements for supervised learning to be an ethical Critical Technical Practice (Agre, 1997). To do so, we will draw on relational ethics theory, as we understand supervised learning to be a relational engagement with a sociotechnical problem space. 

\section{References}
Agre, P. 1997. Toward a Critical Technical Practice: Lesson Learned in Trying to Reform AI. \textit{Social Science, Technical Systems, and Cooperative Work: Bridging the Great Divide}. G.C. Bowker, L. Gasser, S.L. Star, and B. Turner, eds. Erlbaum. 131–158.

Barbosa, N. M., `I\&' Chen, M. (2019, May). Rehumanized crowdsourcing: a labeling framework addressing bias and ethics in machine learning. In \textit{Proceedings of the 2019 CHI Conference on Human Factors in Computing Systems} (pp. 1-12).

Birhane, A. 2021. Algorithmic injustice: a relational ethics approach. \textit{Patterns.} 2, 2 (Feb. 2021), 100205. DOI:https://doi.org/10.1016/j.patter.2021.100205.

Gibney, E. (2020). The battle for ethical AI at the world's biggest machine-learning conference. \textit{Nature}, 577(7791), 609-610.

Greene, D., Hoffmann, A. L., `I\&' Stark, L. (2019). Better, nicer, clearer, fairer: A critical assessment of the movement for ethical artificial intelligence and machine learning. In \textit{Proceedings of the 52nd Hawaii international conference on system sciences.}

Gupta, P., McClatchey, R., `I\&' Caleb-Solly, P. (2020). Tracking changes in user activity from unlabelled smart home sensor data using unsupervised learning methods. \textit{Neural Computing and Applications}, 32(16), 12351-12362.

Mehrabi, N., Morstatter, F., Saxena, N., Lerman, K., `I\&' Galstyan, A. (2021). A survey on bias and fairness in machine learning. \textit{ACM Computing Surveys }(CSUR), 54(6), 1-35.

Shen, H., Deng, W. H., Chattopadhyay, A., Wu, Z. S., Wang, X., `I\&' Zhu, H. (2021, March). Value Cards: An Educational Toolkit for Teaching Social Impacts of Machine Learning through Deliberation. In \textit{Proceedings of the 2021 ACM Conference on Fairness, Accountability, and Transparency} (pp. 850-861).

Sloane, M., Moss, E., Awomolo, O., `I\&' Forlano, L. (2020). Participation is not a design fix for machine learning. \textit{arXiv} preprint arXiv:2007.02423.
\end{document}